\documentclass[prb,aps,twocolumn,amsmath,superscriptaddress]{revtex4}

\usepackage{graphicx}
\usepackage{epsfig}
\usepackage{textcomp,gensymb} 
\usepackage{array}
\newcolumntype{P}[1]{>{\centering\arraybackslash}p{#1}}

\begin{document}
\title{Signatures of four-particle correlations associated with exciton-carrier interactions in coherent spectroscopy on bulk GaAs}

\author{D. Webber}
\affiliation{Department of Physics and Atmospheric Science,
Dalhousie University, Halifax, Nova Scotia B3H 4R2 Canada}

\author{B. L. Wilmer}
\affiliation{Department of Physics and Astronomy,
West Virginia University, Morgantown, WV 26506-6315 USA}

\author{X. Liu}

\affiliation{Department of Physics, University of Notre Dame,
Notre Dame, IN 46556 USA}

\author{M. Dobrowolska}

\affiliation{Department of Physics, University of Notre Dame,
Notre Dame, IN 46556 USA}

\author{J. K. Furdyna}

\affiliation{Department of Physics, University of Notre Dame,
Notre Dame, IN 46556 USA}

\author{A. D. Bristow}
\affiliation{Department of Physics and Astronomy,
West Virginia University, Morgantown, WV 26506-6315 USA}

\author{K. C. Hall}

\affiliation{Department of Physics and Atmospheric Science,
Dalhousie University, Halifax, Nova Scotia B3H 4R2 Canada}

\begin{abstract}

Transient four-wave mixing studies of bulk GaAs under conditions of broad bandwidth excitation of primarily interband transitions have enabled four-particle correlations tied to degenerate (exciton-exciton) and nondegenerate (exciton-carrier) interactions to be studied.  Real two-dimensional Fourier-transform spectroscopy (2DFTS) spectra reveal a complex response at the heavy-hole exciton emission energy that varies with the absorption energy, ranging from dispersive on the diagonal, through absorptive for low-energy interband transitions to dispersive with the opposite sign for interband transitions high above band gap.  Simulations using a multilevel model augmented by many-body effects provide excellent agreement with the 2DFTS experiments and indicate that excitation-induced dephasing (EID) and excitation-induced shift (EIS) affect degenerate and nondegenerate interactions equivalently, with stronger exciton-carrier coupling relative to exciton-exciton coupling by approximately an order of magnitude.  These simulations also indicate that EID effects are three times stronger than EIS in contributing to the coherent response of the semiconductor.          

\end{abstract}


\maketitle
\section{Introduction}
\par{The strong, long-range Coulomb interaction between electrons has a profound influence on the optical response of a variety of material systems, including biological photosynthetic complexes \cite{Engel:2007,Abramavicius:2008}, organic polymer systems \cite{Clark:2007} and semiconductor heterostructures \cite{Shahbook}.   The need to unravel these so-called \textit{many-body} effects, which play a crucial role in a variety of optoelectronic devices such as photodetectors and solar cell technologies, has led to a comprehensive research effort spanning more than two decades\cite{StoneAccChem:2009,AlmandHunter:2014,StoneScience:2009,Cundiff:1996,ElSayed:1997,Allan:1999,Hall:2002,Wang:PRL1993,Shacklette:2002,Borca:2005,Li:2006,Zhang:2007,Karaiskaj:2010,Bristow:2009}.  Semiconductor materials provide an excellent model system for studying Coulomb correlations, in which the interactions between bound electron-hole pairs called excitons may be studied using coherent optical techniques such as time-resolved four-wave mixing (TFWM).  In such experiments, correlations between electron-hole pairs lead to additional contributions to the measured optical signal,\cite{Shahbook} providing a means to separate out the influences of different types of interactions \cite{Cundiff:1996,ElSayed:1997,Allan:1999,Hall:2002,Wang:PRL1993,Shacklette:2002}.  The development of 2DFTS techniques \cite{CundiffPCCP:2014}, in which measurement of the phase of the four-wave mixing signal allows correlations at different absorption and emission frequencies to be identified \cite{Vaughan:2007,BristowRev:2009}, have enabled the further separation of these signal contributions providing additional insight into many-body interactions \cite{StoneAccChem:2009,Borca:2005,Li:2006,Zhang:2007,Bristow:2009,Karaiskaj:2010}.  In recent years, a complex hierarchy of correlations involving successively larger numbers of particles have been revealed using these powerful spectroscopy techniques\cite{Abramavicius:2008,StoneAccChem:2009,AlmandHunter:2014,StoneScience:2009}.}

\par{Existing experiments have primarily focused on exciton-exciton interactions in semiconductor quantum wells \cite{StoneAccChem:2009,StoneScience:2009,Shacklette:2002,Li:2006,Zhang:2007,Karaiskaj:2010,Bristow:2009,Turner:2012,Nardin:2014} in which excitonic resonances involving the heavy-hole and light-hole valence bands are separated in energy but linked to a common ground state via the shared conduction band levels.  Recent studies have revealed bound biexciton states \cite{StoneScience:2009}, enabled the separation of coherent and incoherent contributions to the many-body related signals \cite{Turner:2012}, and have indicated the dominance of effects beyond the Hartree-Fock approximation in the coherent response of the semiconductor \cite{Bristow:2009}.   Much less attention has been paid to nondegenerate four-particle correlations involving excitons and unbound electron-hole pairs \cite{Cundiff:1996,Allan:1999,Hall:2002,Borca:2005,Wehner:1996}, yet for excitation with broad bandwidth laser pulses such effects strongly dominate the overall coherent emission. This was apparent from early TFWM studies that revealed a dramatic enhancement of the exciton response when unbound electron-hole pairs are simultaneously excited \cite{Wehner:1996}.  Coupling of the exciton to unbound electron-hole pairs at higher energies was found to account for this enhancement of the exciton signal \cite{Cundiff:1996,ElSayed:1997}, although the relative importance of EID and EIS was not clear.  2DFTS studies were able to spectrally separate the exciton signal contributions tied to exciton-exciton and exciton-carrier interactions \cite{Borca:2005}, but the relative importance of various coupling effects to the measured signals remained unclear \cite{StoneAccChem:2009,Borca:2005}. } 

\par{Here we report TFWM and 2DFTS experiments on bulk GaAs with the laser pulses energetically tuned above the band gap to elucidate the nature of exciton-continuum interactions.  A strong enhancement of the exciton response in TFWM is observed, and the 2DFTS results indicate separate contributions to the exciton signal tied to degenerate and nondegenerate many-body interactions, consistent with previous reports \cite{Wehner:1996,Allan:1999,Borca:2005}.  The real part of the rephasing 2DFTS signal reveals a rich dispersive structure that was not evident in earlier 2DFTS studies because only the amplitude spectra were detected\cite{Borca:2005,Kuznetsova:2007} or the signal was measured over a narrower range of energies \cite{Zhang:2007,Turner:2012}.  This complex spectral structure prevents a simple assignment of EID and EIS effects to different parts of the spectra, as has been done in previous experiments under resonant excitation of the exciton \cite{StoneAccChem:2009}.   Simulations of the measured 2DFTS results using a density matrix treatment within a multi-level model augmented with a phenomenological treatment of many-body effects provide excellent agreement with both the amplitude and real 2DFTS spectra, including the observed dependence on excited carrier density.  Our simulations indicate that EID and EIS coupling affects degenerate and nondegenerate interactions equivalently, and that exciton-continuum coupling via these effects is an order of magnitude stronger than exciton-exciton coupling, in agreement with past prepulse TFWM experiments\cite{Schultheis:1986}.   For the conditions of our experiments under short pulse excitation of primarily continuum states, four-particle interactions tied to EID have a three-fold stronger influence on the coherent response of the semiconductor than EIS.  }

\begin{figure}[htb]\vspace{0pt}
    \includegraphics[width=8.5cm]{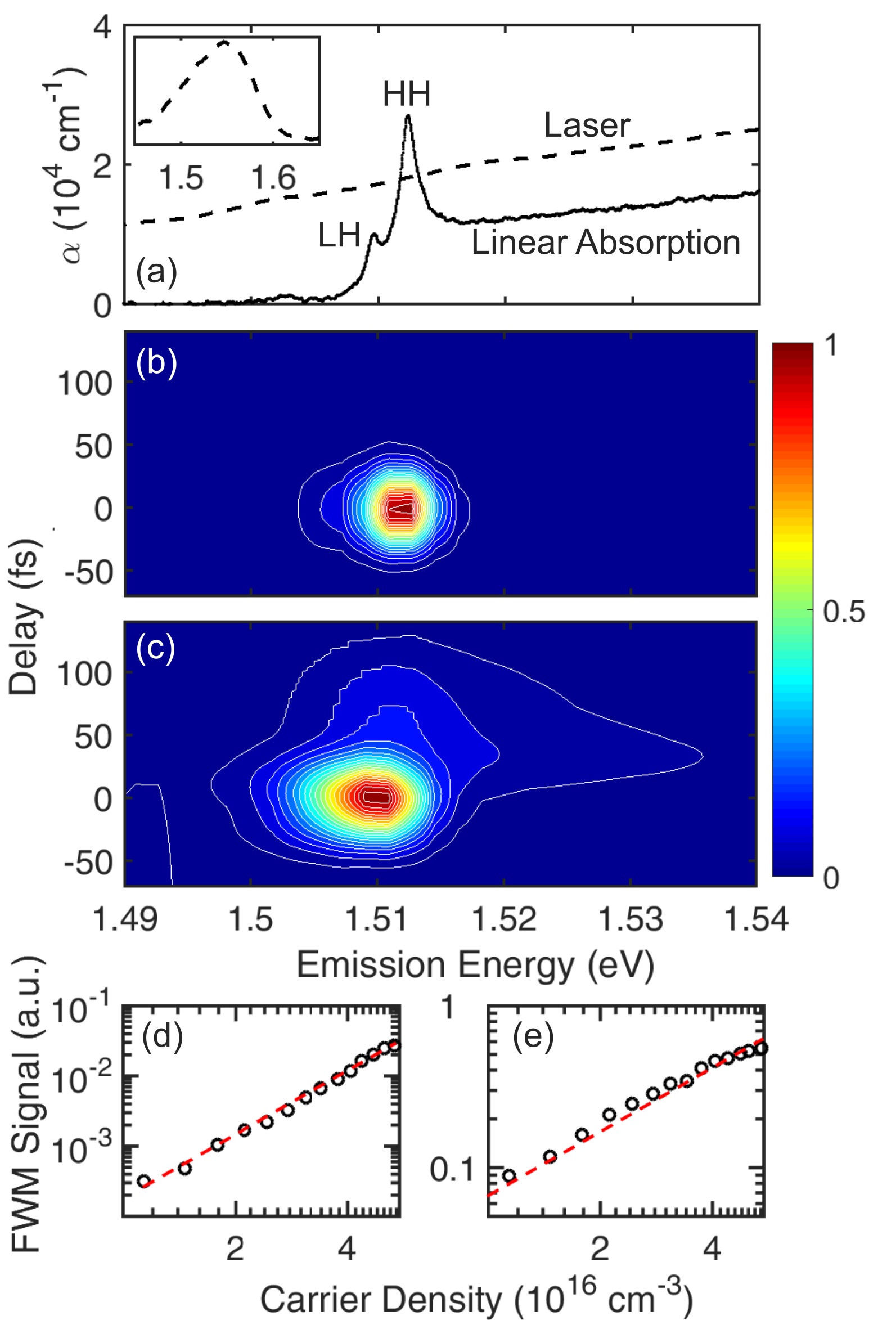}
    \caption{(color online)  (a) Linear absorption (solid curve) of the GaAs sample showing the heavy-hole exciton (HH) and light-hole exciton (LH) resonances together with the laser spectrum (dashed) used in the TFWM experiments. Inset: Laser spectrum plotted over a wider energy range. (b) TFWM results at 10~K as a function of time delay ($\tau$) between $\vec{E_1}$ and $\vec{E_2}$ and detection photon energy for an excited carrier density of $8.0\times 10^{15}$cm$^{-3}$; (c) Same as (b) for a density of $5.4\times 10^{16}$cm$^{-3}$;  (d) The four-wave mixing signal as a function of density for a detection photon energy within the interband continuum at 1.53~eV; (e) Same as (d) for detection at the exciton peak. The red dashed line in (d,e) is a fit to a power law, showing cubic scaling for the interband continuum signal and weaker scaling for the exciton response.}
    \label{fig:Figure1}
\end{figure} 

\section{Materials and Methods}
\subsection{TFWM}
\par{In a TFWM experiment, three optical pulses with wavevectors $\vec{k}_1$, $\vec{k}_2$, and $\vec{k}_3$ are focused onto the sample, generating a third-order nonlinear polarization that emits light in the direction $\vec{k}_3$+$\vec{k}_2$-$\vec{k}_1$. The spectral and temporal composition of this signal contains a wealth of information about the optical transitions within the sample that are in resonance with the laser pulse.  For instance, measurement of the signal as a function of the time delay between the excitation pulses and the detection photon energy may be used to determine the coherence lifetime of excitons and unbound electron-hole pairs\cite{Schultheis:1986,Hugel:1999}, to gain insight into the optical joint density of states of doped semiconductors \cite{Yildirim:2011, Yildirim:2012}, and to study many-body interactions between excited electron-hole pairs \cite{Cundiff:1996, ElSayed:1997,Hall:2002,Webber:2015}.  The high sensitivity of this technique to many-body effects stems from polarization diffraction contributions to the measured signal tied to Coulomb-induced coupling of the polarization on a given optical transition to other transitions \cite{Shahbook}.  TFWM has been used extensively to study the nonlinear optical response of semiconductor systems over the past two decades\cite{Cundiff:1996,ElSayed:1997,Allan:1999,Hall:2002,Wang:PRL1993,Shacklette:2002,Wehner:1996,Rappen:Semicon1994,Webber:2014,Webber:2015,ShackletteJOpt:2003,Yildirim:2011,Yildirim:2012,Schultheis:1986,Hugel:1999,Banyai:1995}.}

\par{Two-pulse, degenerate spectrally-resolved four-wave mixing experiments were carried out on a bulk GaAs sample held at 10~K using 1.55 eV, 30~fs pulses with a full-width at half maximum bandwidth of 78~meV. The laser spectrum is shown in Fig.~\ref{fig:Figure1}(a). Two collinearly-polarized excitation pulses $\vec{E}_1$ and $\vec{E}_2$ with wavevectors $\vec{k}_1$ and $\vec{k}_2$ were focused onto the sample in the self-diffraction geometry (i.e. $\vec{k}_3$=$\vec{k}_2$).  The four-wave mixing signal emitted along (2$\vec{k}_2$-$\vec{k}_1$) was spectrally resolved using a 0.25~m monochromator, and the intensity was measured as a function of the delay between the two excitation pulses using a photomultiplier tube.  The excitation carrier density was estimated using the measured spot size of 70~$\mu$m and by measuring the fraction of transmitted power through the sample from both beams taking into account reflections from the cryostat windows.  Further details regarding the TFWM setup are provided in Ref.~\onlinecite{Webber:2013}.}
\subsection{2DFTS}
\par{Two-dimensional Fourier-transform spectroscopy is an enhanced version of a three-pulse spectrally-resolved four-wave mixing experiment. In contrast to a standard TFWM experiment where the intensity of the four-wave mixing emission is measured, in a 2DFTS experiment both the amplitude and phase of the coherent emission is detected using spectral interferometry techniques \cite{Lepetit:1995}.  By Fourier transforming the coherently detected signal as a function of the time delay between the first and second pulses (the so-called one-quantum rephasing geometry \cite{BristowRev:2009}), one can measure correlations between contributions to the signal at different absorption and emission frequencies.  This enables the separation of signals tied to coupling of resonances via a common ground state \cite{CundiffRevJosaB:2012}, and many-body effects tied to degenerate and nondegenerate interactions \cite{Borca:2005,StoneAccChem:2009}. 2DFTS can also reveal the degree of homogeneous and inhomogeneous broadening of excitonic resonances \cite{KuznetsovaPRB:2007, Bristow:2011}, separate excitonic and biexcitonic contributions to the measured response \cite{StoneScience:2009,Bristow:2009} and detect Raman coherences \cite{CundiffRevJosaB:2012}. Several experimental approaches have been implemented to achieve the necessary phase stability between pulses, including pulse shaping \cite{Vaughan:2007}, diffractive optics \cite{Cowan:2004}, and active interferometric feedback loops utilizing a collinear continuous-wave laser\cite{BristowRev:2009}.} 

\par{Two-dimensional Fourier transform spectroscopy experiments were conducted on the GaAs sample using a multidimensional optical nonlinear spectrometer (MONSTR). More information about this technique, which utilizes three active interferometric feedback loops for phase-stability, can be found in Ref.~\onlinecite{BristowRev:2009}. The process for retrieval of the global phase of the four-wave mixing signal is described in Ref.~\onlinecite{BristowOptExp:2008}. Three collinearly-polarized optical pulses, $\vec{E}_1$,$\vec{E}_2$, and $\vec{E}_3$ with wavevectors $\vec{k}_1$, $\vec{k}_2$, and $\vec{k}_3$ were focused onto the sample that was held at 10~K in a closed cycle optical cryostat. For these experiments, the center photon energy of the laser pulse was 1.527~eV with a bandwidth of 19~meV and a pulse duration of approximately 100~fs. The four-wave mixing signal emitted along $\vec{k}_3$+$\vec{k}_2$-$\vec{k}_1$ was heterodyne detected with a known local oscillator, and the resulting spectral interferogram was measured using a 1~m monochromator. The delay between $\vec{E}_2$ and $\vec{E}_3$ was held fixed at T = 200~fs.}

\subsection{Samples}
\par{The sample investigated in this work was a GaAs epilayer grown by molecular beam epitaxy on a semi-insulating GaAs substrate held at 600$^\circ$C. Prior to deposition of the 800~nm GaAs layer, a 175~nm AlGaAs stop-etch layer was deposited onto the substrate to permit experiments in the transmission geometry. The sample was glued top side down onto a c-cut sapphire window and the substrate was removed using a combination of mechanical polishing and wet-etching.  The linear absorption spectrum of the GaAs sample at 10~K was measured by taking the ratio of the incident and transmitted light using a tunable modelocked Ti:Sapphire oscillator as the excitation source.  The spectra at each laser tuning were detected using a 0.75~m monochromator and photomultiplier tube detector.  The absorption coefficient was extracted from the raw transmission data using a self-consistent model that incorporates multiple reflections within the sample layer. The sample exhibits two peaks at 1.5100~eV and 1.5125~eV, corresponding to the light-hole and heavy-hole exciton resonances, respectively. Due to differing thermal expansion coefficients of GaAs and the sapphire substrate, which were bonded at room temperature, biaxial tensile strain is produced in the GaAs layer at low temperature, lifting the degeneracy of the valence bands \cite{MateoJAP:2007,Wilmer:2016}.}

\section{Numerical Simulations}
\par
\begin{figure}[htb]\vspace{0pt}
    \includegraphics[width=8.5cm]{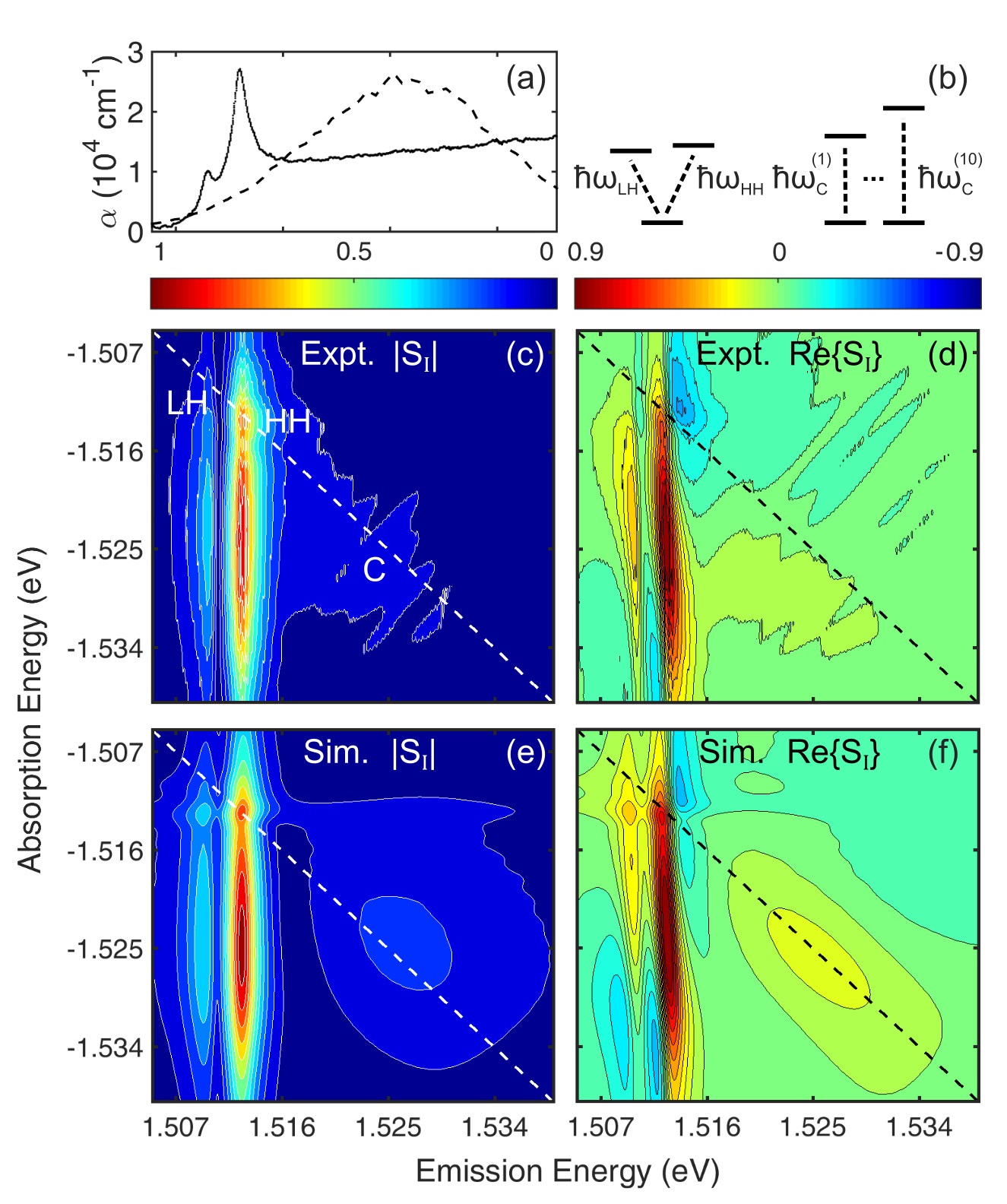}
    \caption{(color online) (a) Linear absorption (solid curve) with the laser spectrum (dashed) used in the 2DFTS experiments. (b) The strain-split heavy-hole and light-hole excitons were modeled as a three-level system with transition energies $\hbar\omega_{HH}$ and $\hbar\omega_{LH}$, respectively. The interband transitions were modeled as ten independent equally-spaced two-level systems with transition energies $\hbar\omega_C^{(1)} .. \hbar\omega_C^{(10)}$. (c)/(d) Experimental results for the amplitude (c) and real part (d) of the 2DFTS signal. (e)/(f) Simulation results for the amplitude (e) and real part (f) of the 2DFTS signal using the optimum EID and EIS parameters.}
    \label{fig:Figure2}
\end{figure}

{We have done numerical simulations of the 2DFTS response of the bulk GaAs sample using a multilevel model within the density matrix approach in the rotating wave approximation.  The multilevel system under study is shown pictorially in Fig.~\ref{fig:Figure2}(b). For the linearly-polarized excitation conditions considered here, the strain-split heavy-hole and light-hole exciton resonances (at energies $\hbar\omega_{\textrm{LH}}$ and $\hbar\omega_{\textrm{HH}}$) form an effective three-level system due to the shared conduction band states. The interband transitions above the band gap are approximated using ten resonances of varying energy ($\hbar\omega_C$), with each resonance separated by 1.8 meV.    The effects of excitation-induced dephasing and excitation-induced shift on the exciton resonances were modeled using a dephasing rate and resonance frequency that depend on the total excited electron-hole population: 
\begin{equation}
\begin{array}{r l}
\frac{1}{T_2} = & \frac{1}{T_{2}^0} + \gamma_{\textrm{EID}}^{\textrm{X}} N_{\textrm{X}} + \gamma_{\textrm{EID}}^\textrm{C} N_{\textrm{C}} \\ 
 \omega = & \omega_0 +  \gamma_{\textrm{EIS}}^{\textrm{X}} N_{\textrm{X}} + \gamma_{\textrm{EIS}}^{\textrm{C}} N_{\textrm{C}} 
\end{array}
\end{equation}
where $T_{2}^0$ and $\omega_0$ are the exciton dephasing time and resonance frequency in the absence of excitation, $N_{X}$ is the sum of the heavy-hole and light-hole exciton densities, $N_{C}$ is the total unbound electron-hole pair density, and $\gamma_{\textrm{EID}/\textrm{EIS}}^{\textrm{X}}$ and $\gamma_{\textrm{EID}/\textrm{EIS}}^{\textrm{C}}$ are the coefficients representing the strength of exciton-exciton and exciton-continuum coupling.   The strength of many-body interactions was assumed to be equal for the heavy-hole and light-hole excitons, and larger for coupling of each exciton transition to the interband continuum than for exciton-exciton coupling to reflect a stronger measured exciton scattering rate involving free carriers \cite{Schultheis:1986}.  This treatment of many-body effects incorporates degenerate and nondegenerate coupling as each excitonic transition is coupled to the heavy-hole and light-hole exciton populations and to population on the full bandwidth of continuum transitions.  A population relaxation time of 15~ps was assumed for all transitions, and a value of $T_{2}^{0}$ of 1 ps was used for the excitonic transitions \cite{Schultheis:1986} and 120 fs for the continuum transitions \cite{Yildirim:2012}.  The many-body coupling coefficients $\gamma_{\textrm{EID}}^{\textrm{C,X}}$ and $\gamma_{\textrm{EIS}}^{\textrm{C,X}}$ were taken as adjustable parameters in fitting the experimental 2DFTS results.  }


\par{In a four-wave mixing experiment, only a single spatial component of the total polarization is measured (e.g. along $\vec{k}_3$+$\vec{k}_2$-$\vec{k}_1$). In the simulation, the four-wave mixing signal was extracted from the polarization through projection of this spatial component\cite{Banyai:1995}. To achieve this, the light field was taken as:
\begin{equation}
\begin{array}{r l}
E(\vec r, t) =& e^{i\vec{k_1}\cdot \vec{r}} \left[ E_0(t) + E_0(t-\tau)e^{i\omega \tau} e^{i\phi} \right.\\
 & \left.+ E_0(t-T)e^{i\omega T}e^{i\psi} \right] + \textrm{c.c.}, 
\end{array}
\label{eqn:ElectricField}
\end{equation}
where c.c. denotes the complex conjugate of the first term, $E_0$ is the pulse envelope (taken to be Gaussian with a pulse width of 100 fs), and $\hbar\omega$ is the energy of the laser pulse (1.527~eV). The light field depends only on two phase factors $\phi = (\vec{k_2}-\vec{k_1})\cdot \vec{r}$ and $\psi = (\vec{k_3}-\vec{k_1})\cdot \vec{r}$. The four-wave mixing signal is then given by:  
\begin{equation}
P(t,\tau) = \frac{1}{4\pi^2}\sum\limits_{x=1}^{4}\sum\limits_{y=1}^{4} P(t,\tau, \phi_x, \psi_y) e^{i\phi_x}e^{i\psi_y}.
\label{eqn:BanyaiThreePulse}
\end{equation}
$P( t,\tau, \phi_x, \psi_y)$ is the macroscopic polarization (determined from the density matrix components) at each delay $\tau$ and each combination of phase angles $\phi_x$ and $\psi_y$. For computational convenience, the phase was taken discretely as $\phi_x = \pi x/2$ and $\psi_y = \pi y/2$. A Fourier transform of the four-wave mixing signal with respect to time and delay yields the emission and absorption frequency axes, respectively.}
%

\section{Results and Discussion}
\subsection{TFWM on Bulk GaAs: Exciton Enhancement Through Nondegenerate Interactions}
\par{Results of spectrally-resolved TFWM experiments for an excitation density of 8.0$\times$10$^{15}$ cm$^{-3}$ are shown in Fig.~\ref{fig:Figure1}(b).  The contour scale indicates the amplitude of the four-wave mixing signal as a function of delay between the two excitation pulses and the emission energy. The laser excitation spectrum for these experiments is shown in Fig.~\ref{fig:Figure1}(a), together with the linear absorption spectrum of the sample.  A single peak is observed at approximately 1.512 eV tied to the combined response of the heavy-hole and light-hole excitons, which are not separately detected due to the limited resolution of the measurements.  The most notable aspect of these results is that the exciton peak strongly dominates the nonlinear optical response of the bulk GaAs sample even though the laser spectrum is tuned above the exciton resonance, exciting primarily the continuum of unbound electron-hole pair transitions.  The exciton response persists only for a narrow range of delay values centered at zero delay, with a full width at half maximum similar to the pulse autocorrelation, despite being spectrally narrow in comparison to the pulse bandwidth.}  

\begin{figure}[htb]\vspace{0pt}
    \includegraphics[width=8.5cm]{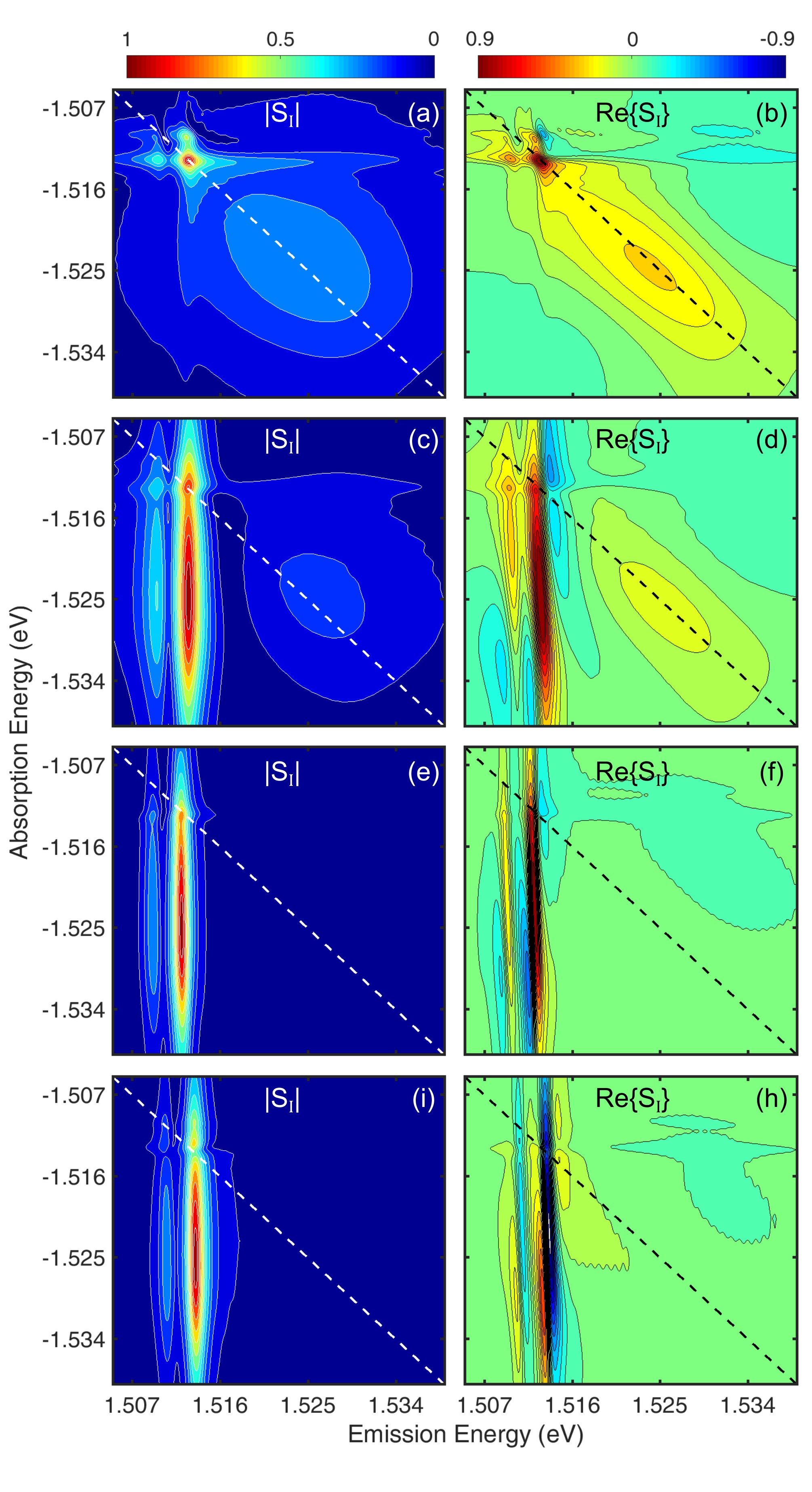}
    \caption{(color online) Simulation results of the normalized 2DFTS signal for varying levels of EID and EIS. (a,b) Amplitude and real part of the calculated 2DFTS signal with no EID or EIS. (c,d) Simulated results with no EIS with EID coefficients governing exciton-exciton and exciton-continuum interactions of $\gamma_{\textrm{EID}}^{\textrm{X}}$=0.27$\times$10$^{-4}$ cm$^{3}$ s$^{-1}$ and $\gamma_{\textrm{EID}}^{\textrm{C}}$=3.6$\times$10$^{-4}$~cm$^{3} $s$^{-1}$, respectively.  (e,f) Simulated results for no EID, but including EIS with the same magnitude of coupling coefficients as in (c,d) with a negative sign. (i,h) Same as (e,f), but with positive values for the EIS coefficients.  }
    \label{fig:Figure3}
\end{figure}

\par{The interband continuum response at emission energies above the exciton resonance is only detectable above the noise floor of the experiments for larger excited carrier densities.  Fig.~\ref{fig:Figure1}(c) shows the results for a density of 5.4$\times$10$^{16}$~cm$^{-3}$.  The continuum response peaks at positive delay values, as expected for a simple inhomogeneously-broadened transition \cite{YT:1979}.    The continuum response scales as the cube of the intensity over several orders of magnitude [(Fig.~\ref{fig:Figure1}(d)], verifying that the system response is within the $\chi^{(3)}$ regime at these excitation densities.  A sub-cubic scaling at the exciton is expected for all densities due to the influence of many-body effects,\cite{Allan:1999,ShackletteJOpt:2003} in line with the results in Fig.~\ref{fig:Figure1}(e). The general features of these results are consistent with earlier studies using TFWM in GaAs and other bulk and quantum well semiconductor systems\cite{Cundiff:1996,Allan:1999,Hall:2002,Wehner:1996,Rappen:Semicon1994,Webber:2014,Webber:2015}.} 

\begin{figure*}[htb]\vspace{0pt}
    \includegraphics[width=17cm]{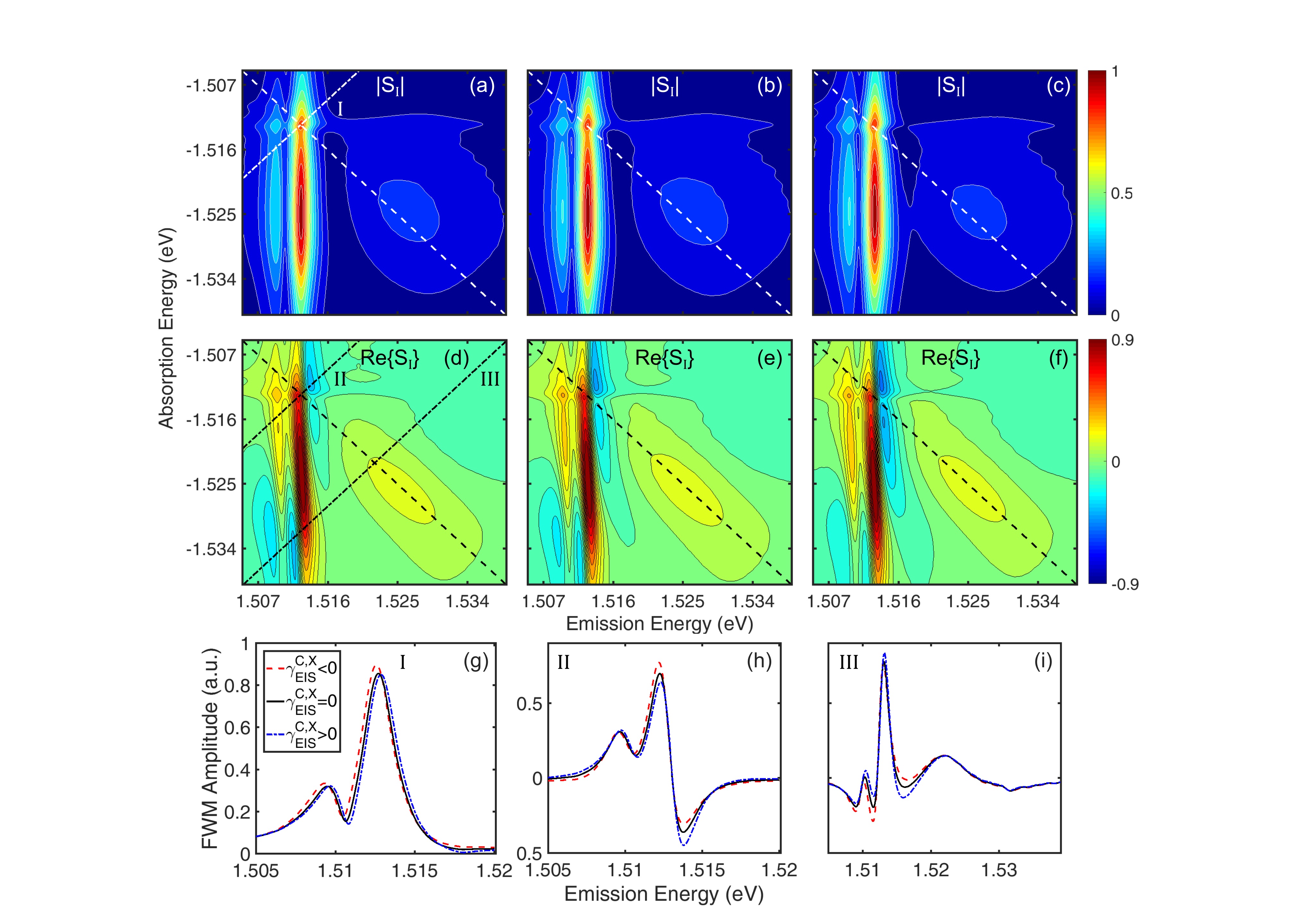}
    \caption{(color online) Simulation results of the normalized 2DFTS signal for varying levels of EIS with the EID coefficients held constant at the optimum values. (a) Simulated results with negative EIS coefficients three times smaller in magnitude than the EID coefficients; (b) Simulated results without EIS added.  (c) Same as (a) with positive EIS coefficients.  (d-f) Real part of 2DFTS signal shown in (a-c). Cross diagonal slices of the amplitude and real 2DFTS spectra are plotted along lines I, II, and III as a function of emission energy in (g), (h), and (i) respectively. }
    \label{fig:Figure4}
\end{figure*}

\par{The observed exciton enhancement in TFWM under broad bandwidth excitation is caused by nondegenerate coupling of the exciton to the unbound electron-hole pair population, which leads to contributions to the measured signal emitted at the exciton energy induced by absorption at energies within the interband continuum.\cite{ElSayed:1997} Due to the inability to separate signals tied to absorption at different energies in TFWM, this physical understanding of the observed four-wave mixing response only emerged through detailed theoretical calculations\cite{ElSayed:1997}, and two-color TFWM experiments\cite{Cundiff:1996}. The associated exciton signal has been referred to as the \emph{continuum contribution} (CC) \cite{ElSayed:1997}, and is larger the broader the bandwidth of continuum transitions excited.  The short duration of the exciton response versus the delay between the two excitation pulses is caused by interference of signal contributions involving interband excitations at different energies, which is constructive only in the vicinity of zero delay where a net grating in the total free carrier population summed over energy exists \cite{ElSayed:1997,Cundiff:1996}.}  

\par{The type of exciton-continuum coupling included in the original treatments of the CC signal at the exciton was EID \cite{ElSayed:1997}, however the role of EIS was highlighted in later experiments in semiconductor quantum wells \cite{Shacklette:2002}.   EID and EIS are both caused by four-particle correlations that lead to renormalization of the exciton self-energy, with EIS (EID) tied to the real part (imaginary part) \cite{StoneAccChem:2009,Zhang:2007}.  (While local field contributions tied to polarization-polarization interactions may also contribute, the associated four-wave mixing signal was found to be weaker than the EID/EIS contributions for the broadband excitation conditions considered here \cite{Rappen:Semicon1994}.)  For the case of nondegenerate coupling, the four-particle correlations are between bound excitons and unbound electron-hole pairs excited on the continuum of interband transitions.  The CC signals tied to EID and EIS will both be emitted at the exciton energy and persist for a narrow range of pulse delays around zero delay (dictated by the existence of a net population grating of free carriers).   As a result, it is not possible to disentangle the contributions of EID and EIS in TFWM.  Since EID and EIS contributions to the four-wave mixing polarization are 90 degrees out of phase\cite{Li:2006}, studies of the exciton response using 2DFTS techniques provide a means of elucidating the relative importance of these two coupling processes, as discussed further below.}

\subsection{2DFTS Results: Separation of Signals Tied to Exciton-Exciton and Exciton-Continuum Coupling}
\par{The results of 2DFTS experiments conducted on the bulk GaAs sample are shown in Fig.~2(c) and Fig.~2(d) for the normalized amplitude and real part of the rephasing four-wave mixing response (S$_{\textrm{I}}$), respectively. These results correspond to an excited carrier density of 1.6$\times$10$^{16}$ cm$^{-3}$.  As in the TFWM experiments, the laser energy is tuned above the exciton resonance to excite primarily continuum transitions, as shown in Fig.~\ref{fig:Figure2}(a).  The amplitude 2DFTS spectrum in Fig.~\ref{fig:Figure2}(c) is composed of two primary features:  (i) vertical stripes at emission energies of 1.5100~eV and 1.5125~eV, corresponding to the light-hole and heavy-hole exciton resonances; and (ii) a broad-band response at energies above the exciton resonances that resides along the diagonal axis.  The heavy-hole exciton response is stronger than the light-hole response due to the weaker oscillator strength of the light-hole transitions and the greater degree of overlap of the laser pulse spectrum with the HH exciton resonance. The broad band response is due to the continuum of interband transitions, with a signal that covers an energy range similar to the laser spectrum.} 

\begin{figure*}[htb]\vspace{0pt}
    \includegraphics[width=17cm]{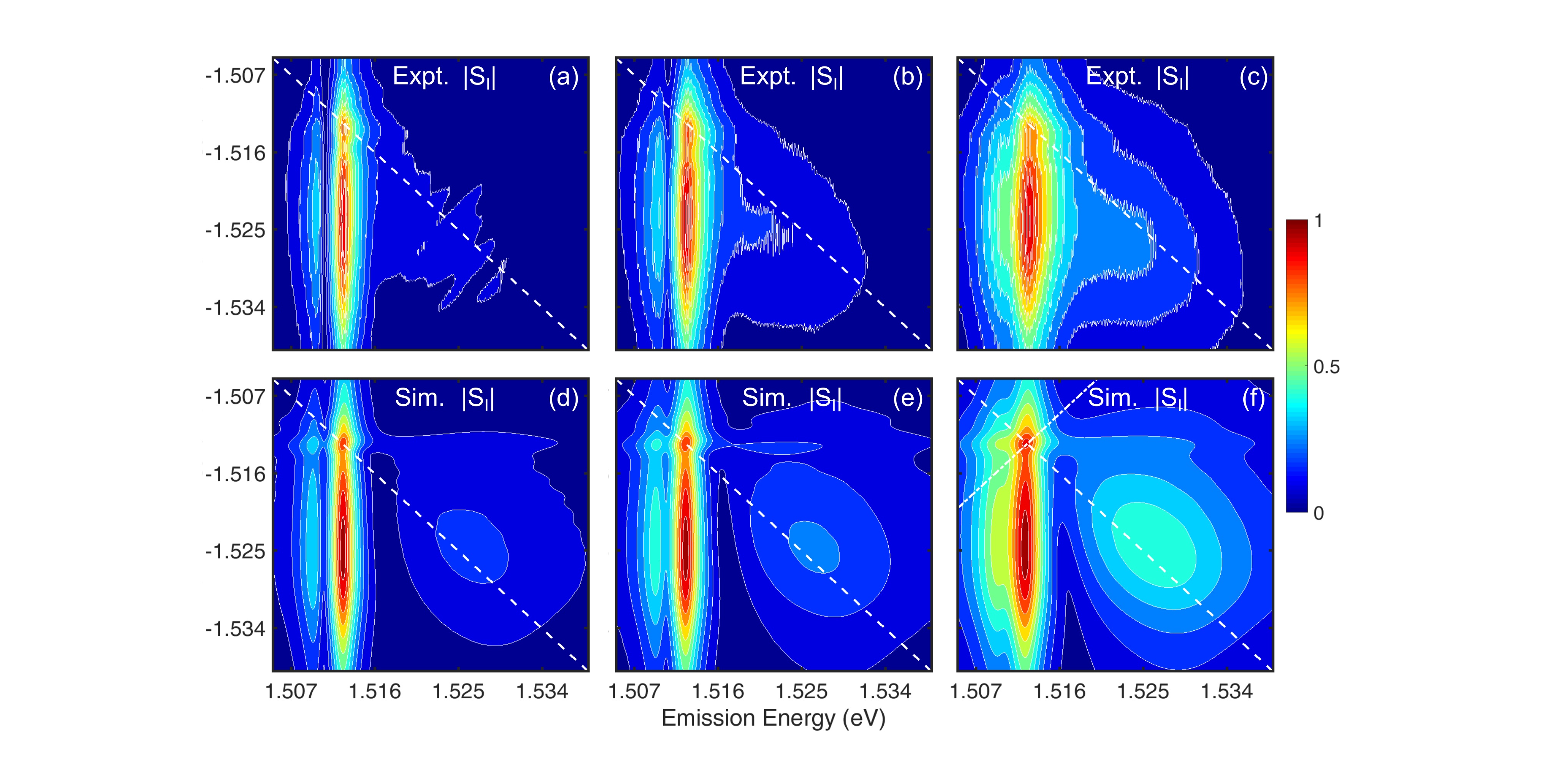}
    \caption{(color online) Normalized amplitude results of 2DFTS experiments and simulations for varying excited carrier density. (a-c) Results of 2DFTS experiments for excited carrier densities of $1.6\times 10^{16}$ cm$^{-3}$, $3.2\times 10^{16}$ cm$^{-3}$, and $8.0\times 10^{16}$ cm$^{-3}$ respectively. (d-f) Simulation results for the same excited carrier densities as (a-c).}
    \label{fig:Figure5}
\end{figure*}

\par{The two vertical stripes corresponding to emission at the light-hole and heavy-hole exciton resonances consist of a number of spectrally distinct contributions along the absorption axis.  There are peaks along the diagonal tied to resonant absorption and emission at the exciton resonances as well as discrete off-diagonal peaks tied to coupling between the heavy-hole and light-hole excitons.  The latter coupling effects have been studied extensively using 2DFTS techniques\cite{Li:2006,Zhang:2007,Karaiskaj:2010,Bristow:2009}.  The vertical stripes at larger absorption energies are caused by absorption within the interband continuum followed by emission at the exciton resonances.  These signals correspond to the CC that leads to strong enhancement of the excitons in TFWM experiments when the continuum of interband transitions is excited together with the excitons, as discussed in Sec.~IIIA.   The ability to separate absorption and emission pathways in 2DFTS permits the separation of the discrete resonance contributions and the CC. Based on the observed amplitudes, the CC strongly dominates the overall four-wave mixing response for conditions of broad bandwidth excitation of primarily continuum transitions, a result that is consistent with conclusions based on prior TFWM experiments \cite{Cundiff:1996,Allan:1999,Hall:2002,Wehner:1996,Rappen:Semicon1994,Webber:2014,Webber:2015}.}

\par{The real part of the 2DFTS spectrum in Fig.~\ref{fig:Figure2}(d) reveals a complex spectral structure that varies along the absorption energy axis.  Focusing on the signal emitted at the HH exciton resonance, the spectral shape on the diagonal is dispersive in nature, in agreement with previous work,\cite{Li:2006,Zhang:2007,Turner:2012} while for higher absorption energies the lineshape evolves from a primarily absorptive response to a dispersive structure as the absorption energy increases.  The dispersive structure at high energies within the continuum is 180$^{\circ}$ out of phase with the exciton response on the diagonal.  The vertical stripe feature associated with the CC in Fig.~\ref{fig:Figure2}(c) and Fig.~\ref{fig:Figure2}(d) was observed previously in rephasing 2DFTS experiments but the variation in the spectral structure for absorption energies within the continuum in Fig.~\ref{fig:Figure2}(d) was not evident either because only the amplitude spectrum was detected\cite{Borca:2005,Kuznetsova:2007} or the signal was measured over a narrower range of energies \cite{Zhang:2007,Turner:2012}.  Unraveling the spectral structure within the CC in Fig.~\ref{fig:Figure2}(d) would provide insight into exciton-continuum interactions within the optically excited semiconductor.  }

\renewcommand{\arraystretch}{1.5}
\begin{table}[]
\centering
\caption{Many-body coefficients ($\gamma_{\textrm{EID/EIS}}^{\textrm{X}}$ and $\gamma_{\textrm{EID/EIS}}^{\textrm{C}}$) found in numerical simulations of the 2DFTS results to provide the best agreement with the experimental data.}
\label{my-label}
\begin{tabular}{P{2.8cm} P{2.8cm} P{2.8cm}}
\hline
\underline{Coefficients} & \underline{Exciton-Exciton} & \underline{Exciton-Continuum}   \\ \hline
$\gamma_{\textrm{EID}}$ (cm$^{3}$~s$^{-1}$) & 0.27$\times 10^{-4}$  & 3.6$\times 10^{-4}$   \\
$\gamma_{\textrm{EIS}}$ (cm$^{3}$~s$^{-1}$) & 0.09$\times 10^{-4}$  & 1.2$\times 10^{-4}$ \\
\hline
\end{tabular}
\end{table}

\par{The different lineshape of the HH exciton emission on the diagonal and for higher absorption energies makes it tempting to conclude that the primary coupling mechanism influencing the HH exciton emission differs for degenerate and nondegenerate interactions.  In two-quantum studies on GaAs quantum wells \cite{StoneAccChem:2009}, the dispersive structure at the exciton was attributed to EIS and the absorptive shape in the continuum was attributed to EID since these two types of interactions within a simple two-level system analysis are known to generate polarization responses that are 90$^{\circ}$ out of phase \cite{Shacklette:2002}. The variation in spectral shape for varying absorption energy in Fig.~\ref{fig:Figure2}(d) makes such conclusions more difficult to draw.  While the dominance of the CC contribution to the overall four-wave mixing response of the GaAs sample is clear, supporting numerical simulations are needed to identify the relevant coupling mechanisms within the system of bound and unbound electron-hole pairs.}

\subsection{Simulations of 2DFTS Experimental Results}
\par{Simulations of the 2DFTS results were carried out for varying strengths of EID and EIS.  The simulation results corresponding to the optimum coupling parameters are shown in Fig.~\ref{fig:Figure2}(e) and Fig.~\ref{fig:Figure2}(f) for the amplitude and real 2DFTS spectra, respectively.   The evolution of the dispersive structure at the HH exciton detection energy with increasing absorption energy is well reproduced  in the simulated results.  The optimum many-body coefficients tied to EID were found to be $\gamma_{\textrm{EID}}^{\textrm{X}}$=0.27$\times$10$^{-4}$ cm$^{3}$ s$^{-1}$ and $\gamma_{\textrm{EID}}^{\textrm{C}}$=3.6$\times$10$^{-4}$~cm$^{3} $s$^{-1}$.  These values indicate an approximately ten-fold larger rate of exciton-continuum scattering than exciton-exciton scattering, consistent with the experimental results in Ref.~\onlinecite{Schultheis:1986}.  The optimum values of the EIS coefficients were found to be smaller than the EID coefficients by the same ratio for the exciton-exciton and exciton-continuum interactions, corresponding to a factor of approximately three.  EIS therefore plays a smaller role than EID in the coherent response of the exciton. It is notable that good agreement is obtained for the full range of emission and detection energies using the same relative strengths of EID and EIS describing both degenerate and nondegenerate interactions, indicating that exciton-exciton and exciton-continuum interactions are governed by the same coupling mechanisms.  }

\par{In order to gain insight into the spectral dependence of the signal contributions tied to EID and EIS, simulations were carried out with EID or EIS alone [Fig.~\ref{fig:Figure3}] or by varying the level of EIS in the presence of EID held fixed at the optimum value [Fig.~\ref{fig:Figure4}].  The vertical stripe associated with the CC is observed in the calculated amplitude 2DFTS results with a similar spectral shape for any combination of EID and/or EIS, but the real 2DFTS spectra vary considerably in their qualitative features as the relative amounts of the two physical effects are varied.  At the exciton resonance on the diagonal, EID alone is sufficient to recover the dispersive structure observed in many past experiments \cite{Li:2006,Zhang:2007,Turner:2012}.  Simulations incorporating a similar phenomenological treatment of EID and EIS were carried out in Ref~\onlinecite{Li:2006} considering only the discrete exciton transitions, which were excited resonantly in that work.  They found a dispersive feature was only produced at the exciton when EIS was included.  The simulations in Fig.~\ref{fig:Figure3} and Fig.~\ref{fig:Figure4} show that a simple separation of signal characteristics tied to EID and EIS is not possible under nonresonant excitation conditions, for which continuum states are excited together with the exciton.}  

\par{The spectral structure of the calculated 2DFTS response with varying absorption energies agrees qualitatively with the measured results when EID alone is included in the simulations.  For the signal emitted at the heavy-hole exciton, EID reproduces both the correct sign of dispersion and the relative degrees of dispersion (\textit{i. e.} the magnitude of the negative features relative to the positive ones) for absorption at the exciton and within the interband continuum.  Changes in the magnitude of the EID coefficient affect the overall linewidth of the exciton resonance, and the optimum values provide good agreement with the experimental exciton linewidth.   For a positive value of the EIS coefficient and no EID included [Fig.~\ref{fig:Figure3}(h)], the calculated spectra are in very poor agreement with the experimental results.  When only EIS is included with a negative sign for the EIS coupling coefficient [Fig.~\ref{fig:Figure3}(f)], the qualitative shape of the spectra is correct, but the line width of the exciton resonance is smaller than in the experiment and the relative magnitudes of the negative dip at the exciton and continuum absorption energies are reversed.  }
  

\par{The agreement between the measured results and the calculations including EID improves with the addition of a small amount of EIS with a negative sign in the coefficients.  The variations in the signal characteristics with changes in the magnitude and sign of the EIS coefficients with the EID coefficients held at the optimum values are highlighted in Fig.~\ref{fig:Figure4} through cross diagonal cuts at the exciton resonance in the amplitude 2DFTS response [Fig.~\ref{fig:Figure4}(g)], at the exciton resonance in the real 2DFTS response [Fig.~\ref{fig:Figure4}(h)], and at absorption energies corresponding to the CC response in the real 2DFTS results [Fig.~\ref{fig:Figure4}(i)].  The cross diagonal cut at the exciton in Fig.~\ref{fig:Figure4}(g) shows that a positive (negative) value of the EIS coefficient shifts the HH exciton resonance to higher (lower) energies along the emission axis.   In the real 2DFTS spectra, EIS has the effect of increasing or decreasing the amount of asymmetry in the dispersive peak at the HH exciton emission energy.  In the experimental results, the HH exciton resonance peak is centered on the diagonal in the amplitude 2DFTS response.  The simulations produce a centered exciton peak for negative EIS coupling coefficients three times smaller than the corresponding EID coefficients, reflecting the dominant role played by EID in the coherent response of the semiconductor.}

\section{Dependence on Excited Carrier Density}
\par{The dependence of the four-wave mixing signal on excited carrier density is shown in Fig.~\ref{fig:Figure5}. The amplitude 2DFTS results for densities of 1.6$\times 10^{16}$, 3.2$\times 10^{16}$, and 8.0$\times 10^{16}$~cm$^{-3}$ are shown in Fig.~\ref{fig:Figure5}(a)~-~Fig.~\ref{fig:Figure5}(c), respectively. The dominant effect of increasing the carrier density is an increase of the linewidth of the exciton along the emission axis, reflecting the strong role played by EID in the exciton response.  At the highest carrier density probed, the collision-induced broadening is sufficient that the energetically split heavy-hole and light-hole exciton peaks merge. In addition to the observed broadening, the strength of the interband continuum response on the diagonal is enhanced relative to the exciton four-wave mixing signal. The slower growth of the exciton signal with increasing carrier density relative to the interband response is also caused by EID.  As discussed previously \cite{Allan:1999,ShackletteJOpt:2003} this sub-cubic scaling behavior of the exciton is caused by the larger total dephasing rate of the exciton at higher densities, which reduces the time-integrated signal.  The simulated 2DFTS results for the optimum EID and EIS coupling parameters provide good agreement with the measured carrier density dependence, as shown in Fig.~\ref{fig:Figure5}(d)-Fig.~\ref{fig:Figure5}(f).  Since the EID coefficient is held constant in the simulations, the good agreement suggests that screening of the EID coefficient is negligible at the carrier densities considered here.}

\section{Conclusions}
\par{In summary, TFWM and 2DFTS experiments were carried out on bulk GaAs to elucidate the nature of exciton-carrier interactions.  TFWM results for broadband excitation of primarily interband transitions indicated a strong enhancement of the exciton emission in TFWM, a result that was shown using 2DFTS to result from nondegenerate four-particle correlations between bound and unbound electron-hole pairs.  Real 2DFTS spectra revealed a complex dispersive structure:  For emission at the heavy-hole exciton, the discrete response on the diagonal was primarily dispersive, in agreement with previous work, but for absorption within the interband continuum, the response at low-energies was primarily absorptive and at high energies dispersive with a 180$^{\circ}$ phase shift relative to absorption at the exciton.  Despite these differences, simulations based on a simple multilevel model augmented by many-body effects provide excellent agreement with the experiments and show that the same physical effects govern exciton-exciton and exciton-continuum interactions.  We find that the spectral features are qualitatively captured considering coupling via excitation-induced dephasing alone, with a three-fold weaker EIS included to enhance the quantitative agreement.  The EID coupling parameters governing exciton-exciton and exciton-continuum interactions determined from fitting the experimental 2DFTS spectra are in line with values found in past TFWM experiments.  }

\par{Our findings highlight the importance of simulations in determining the operative many-body effects when continuum transitions are excited together with the exciton, as a simple assessment of the relative role of different coupling mechanisms is not possible using a discrete resonance analysis and identification of the measured lineshapes.  While the excellent agreement obtained using the phenomenological treatment of many-body effects used here illustrates the power of these simple approaches for determining the dominant interactions, a full many-body calculation using a microscopic model such as dynamics-controlled truncation \cite{Axt:1994,Lindberg:1994} would aid in the interpretation of the effects observed.   In addition, extension of the 2DFTS experiments reported here to larger waiting times\cite{Turner:2012} would provide useful verification of the interpretation presented here, in which only exciton renormalization tied to population terms were needed to account for the experimental observations.  Our findings provide new insight into the role of many-body interactions in the coherent optical response of semiconductors. }

\par{The work performed at Dalhousie University was supported by the Natural Sciences and Engineering Research Council
of Canada and the Walter C. Sumner Foundation. Work at West Virginia University was supported by the West Virginia Higher Education Policy Commission (HEPC.dsr.12.29) and the National Science Foundation (CBET-1233795).  Work at the University of Notre Dame was supported by the National
Science Foundation (Grant DMR1400432).}

\end{document}